\documentclass{article}
\usepackage{graphicx}
\usepackage{authblk}

\newcommand{\NI}{\noindent}

\begin{document}

\title{An Upper Bound on the Number of Discrete States Possible for the Human Brain}
\author{Jon Borresen$^1$, Killian O'Brien}
\date{$^1$J.Borresen@mmu.ac.uk \\ \vskip 5mm \today}
\affil{School of Mathematics and Computation, Manchester Metropolitan University, UK}

\maketitle

\begin{abstract}
Human brains are arguably the most complex entities known.  Composed of billions of neurons, connected via a highly detailed structure where the underlying method by which functionality occurs is still debated.  Here we consider one theory for neural coding, synchronization coding, which gives rise to the highest possible number of discrete states that a brain could exist in. A strict upper bound on the number of these states is determined.  We conclude that the theoretical upper limit on the capacity of one human brain is almost inconceivably large and massively larger than the corresponding theoretical limit that could be obtained using every transistor ever built.  

\end{abstract}
\section{Introduction}
In $2011$, the journal Science \cite{Hilbert} reported that the total computing power of the world was approximately equal to that of one human brain.  Since then, the increase in total computing power has been considerable and if we accept that Moore's Law is valid \cite{Moore} (and the implications thereon), it would seem that our humble biological devices are quickly becoming obsolete.\\

The original article used ``the maximum number of nerve impulses executed by one human brain per second." as a measure of computing power.  If we take an average human brain of $8.6 \times 10^{12}$ spiking neurons \cite{Azevedo}, firing at a maximum frequency of $300$ Hertz, we arrive at an estimate of the human brain's computing power of $2.58 \times 10^{15}$ operations per second.  This is less than the estimate in Science, which also considered the number of connections between each neuron (although it is difficult to equate connectivity to computational power in any direct way).\footnote{We are not using the standard floating point operations per second (FLOPS) as this infers too much about how brains are operating.} If we are considering the number of discrete states a brain could exist in if neurons operated in a simple binary on/off manner (I.E. each neuron were performing in a manner similar to a transistor), we obtain the total number to be approximately $2^{8.6\times 10^{12}}\approx 10^{258,885,796,271}$ (bits) - which would seem a fairly impressive memory capacity and massively larger than the $2011$ estimate for the total memory power of all the computers in the world at $2.36 \times 10^{21}$ bits \cite{Hilbert}.\\

However, brains are not simple binary computing devices and operate in a very different manner to standard computers.  The fundamental mechanisms by which brains process and store information may give rise to higher or lower numbers of operations to that stated above.\\

Here we ask a hypothetical question: \textit{``Given the various theories of neural coding, what is the theoretical upper bound on the computational capacity of the human brain?"}  This is in many ways akin to asking ``What is the lifespan of the universe?'' and concluding this as the theoretical upper bound on ``How long will I live?''  As such, the answer to the first question gives little information as to the answer to the second but is nonetheless a valid and interesting question in its own right.  I.e. We are not concerned with how many discrete dynamical states the human brain can actually exist in, but what is the theoretical upper bound on this.\\

Clearly the number of discrete states cannot be infinite.  If it were possible to store an infinite number of bits in a brain of 8 billion neurons, it would be also possible to do the same with half that number, and half again - there would be no need for a big human sized brain and we would all have much smaller heads.\\

\section{The Human Brain}

The human brain is massively complex and it is beyond the scope of this article to fully describe how it operates.  It is useful though to have a basic idea of what we are considering, if only to frame the concept we are trying to investigate.  \\

From a functional perspective the brain is composed of an enormous number of cells, connected via an extremely complex network.  These cells are, in the majority, glial cells, which provide the physical structure of the brain and are involved in removal of waste and other non-information processing functions.  About $10\%$ of brain cells are neurons, which are the fundamental entities that perform the processing and memory.\\

Neurons are themselves very complex entities.  Long, thin and able to form multiple branches (dendrites). They use electrochemical pulses to transmit information to other neurons.  Each neuron connects to on average $1000$ other neurons \cite{Williams} via a synapse - a small gap between neurons across which neurotransmitters diffuse.  In turn, the neurotransmitters can either polarize or depolarize the neuron to which it is connected - causing the post-synaptic neuron to pulse or not to pulse.\\

It is possible to model the pulsating behaviour of neurons using systems of ordinary differential equations.  The most famous of these being the Hodgkin-Huxley model \cite{Hodgkin}, which accurately simulates the electrochemical pulse moving down the axon of a neuron.  This model is highly detailed, having differential equations to describe gating channels as well as the voltage.  The equations are somewhat tricky to numerically integrate and the number of underlying parameters is large.\\

A simpler and more accessible neuron model is the Fitzhugh-Nagumo model \cite{Fitzhugh,Nagumo}, which is essentially a reduction of the Hodgkin-Huxley model described above.  The governing equations are given as:

\begin{eqnarray} \label{eqn:FHN}
\dot{u}&=&c(-v+u-u^3/3+I)\nonumber \\
\dot{v}&=&u-bv+a,
\end{eqnarray}

\NI where $I$ is some external current applied to the neuron, $a, \, b$ and $c$ are the neuron's parameters and the variables $u$ and $v$ correspond to the fast (spiking) and slow voltages.\\

\begin{figure}
\begin{center}
\includegraphics*[height=40mm, width=70mm]{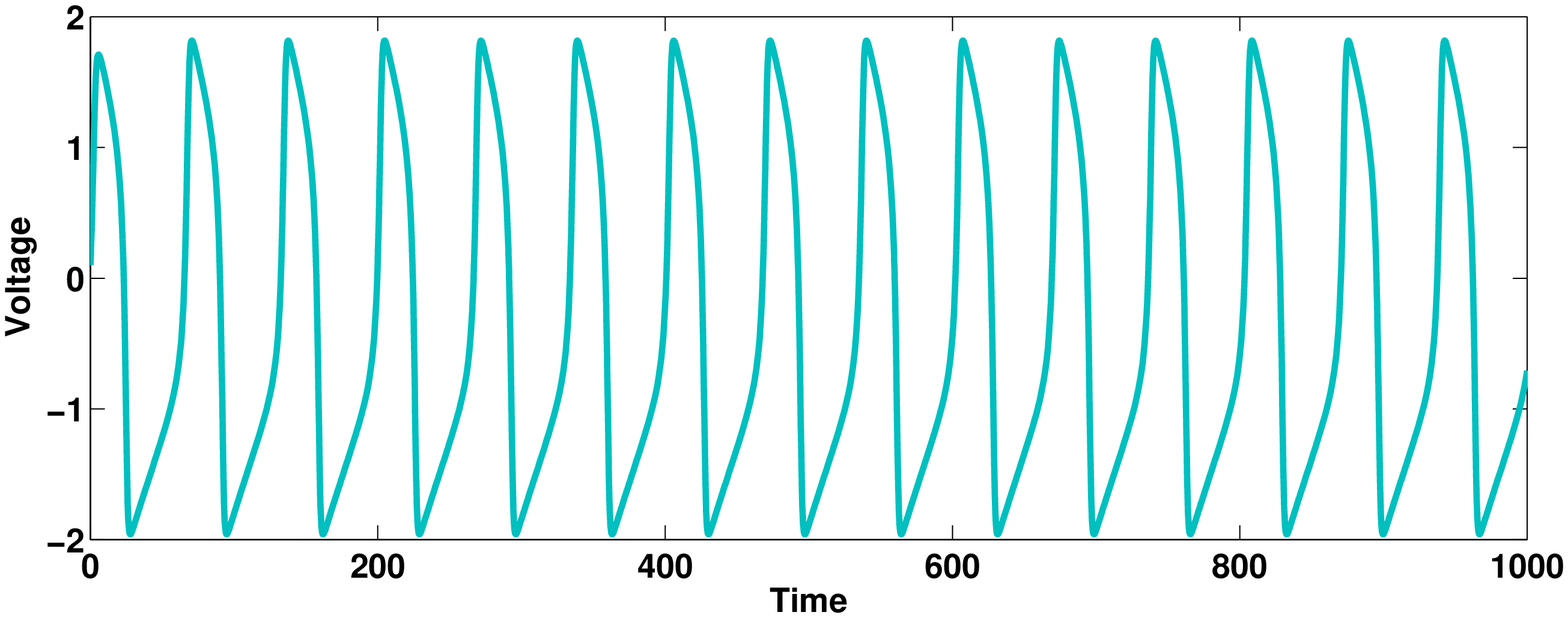}
\caption{\label{fig:FN1} {\bf{Fast Voltage of Fitzhugh Nagumo Oscillator}}  The parameter values are: $a=0.7;b=0.8;c=10;I=0.5$.}
\end{center}
\end{figure}

\section{Theories of Neural Coding}

There is still some debate as to how brains store and process information.  There are various theories of neural coding and it is generally believed that more than one fundamental mechanism is used.  As a summary, the following classification would cover most of the available theories:

\begin{itemize}
\item Population
\item Rate
\begin{itemize}
\item Spike Count
\item Time Dependent Firing Rate
\end{itemize}
\item Spatio-Temporal
\begin{itemize}
\item Binary
\item Receptive Field (this generally applies only to the retina)
\item Synchronization
\end{itemize}
\end{itemize}

If we are concerned with determining an upper bound on the computational power we would need only consider the coding mechanism which gives the highest theoretical number of states, in this case, synchronization coding.\\

\section{Synchronization Coding}
Synchronization coding is a form of spatio-temporal coding in which information is stored not in the individual firing of neurons but in the similar response to stimulus of groups of neurons. \\

The Fitzhugh-Nagumo Model (Equation \ref{eqn:FHN} above) can be adapted to demonstrate synchronization similar to that observed in neurons via coupling through the fast gating variable $u$. For a population of $n$ neurons the governing equations are:
\begin{eqnarray} \label{eqn:FHN2}
\dot{u_i}&=&c(-v_i+u_i-u_i^3/3+I)+\sum_{j=1}^n k_{i,j}u_j\nonumber \\
\dot{v_i}&=&u_i-bv_i+a_i,
\end{eqnarray}

where $k_{i,j}$ represents the coupling strength between neuron $i$ and $j$. For $k>0$ we tend to observe synchronization between neuron $i$ and $j$ and for $k<0$ the neurons tend to desynchronize.\\

Although we present here a very simplified form of coupling we are in essence, retaining the underlying neural dynamics of excitation and inhibition observed in biology.\\

\begin{figure}
\begin{center}
\includegraphics*[height=40mm, width=70mm]{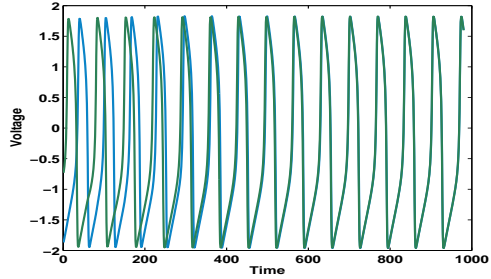}
\caption{\label{fig:FN2Sync} {\bf{Synchronization in 2 Coupled Fitzhugh Nagumo Equations:}}  The parameter values are: $a=0.7;b=0.8;c=10;I=0.5;$  The equations are coupled through the fast gating variable ($u$) with coupling strength $0.01$.  From non-identical initial conditions the neurons quickly synchronize.}
\end{center}
\end{figure}

Figure \ref{fig:FN2Sync} demonstrates synchronization in two coupled Fitzhugh-Nagumo neurons.  Although beginning with different dynamics they rapidly synchronize.  A perturbation to one neuron will in turn affect the other and synchronization would be restored.\\

We can, by selecting suitable coupling strengths, cause larger populations of neurons to form into groups (known as clusters) performing similar actions. A variety of cluster states can be achieved using varying coupling strengths between the neurons. \\ 

\begin{figure}
\begin{center}
\includegraphics*[height=40mm, width=70mm]{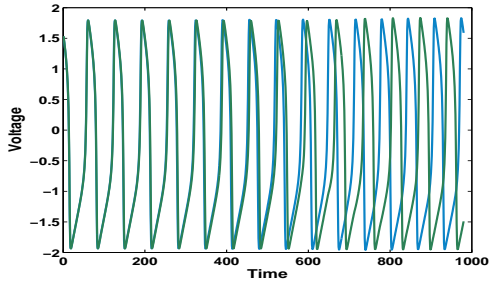}
\caption{\label{fig:FN2Desync} {\bf{Desynchronization in 2 Coupled Fitzhugh Nagumo Equations:}}(Parameter values as in \ref{fig:FN2Sync} with $k=-0.1$).  From identical initial conditions the oscillators quickly desynchronize to give alternating spikes.}
\end{center}
\end{figure}

\begin{figure}
\begin{center}
\includegraphics*[height=40mm, width=70mm]{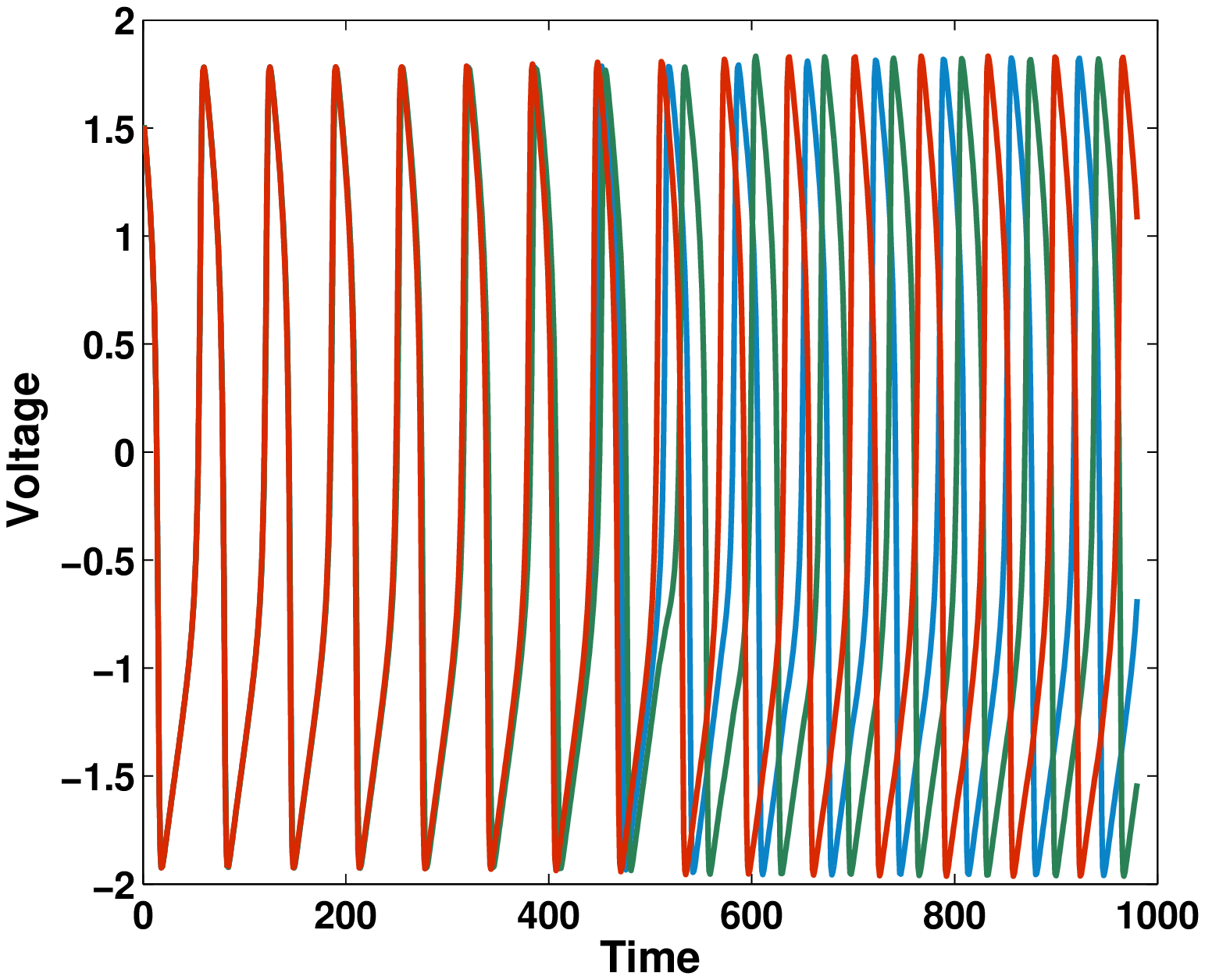}~~
\includegraphics*[height=40mm, width=70mm]{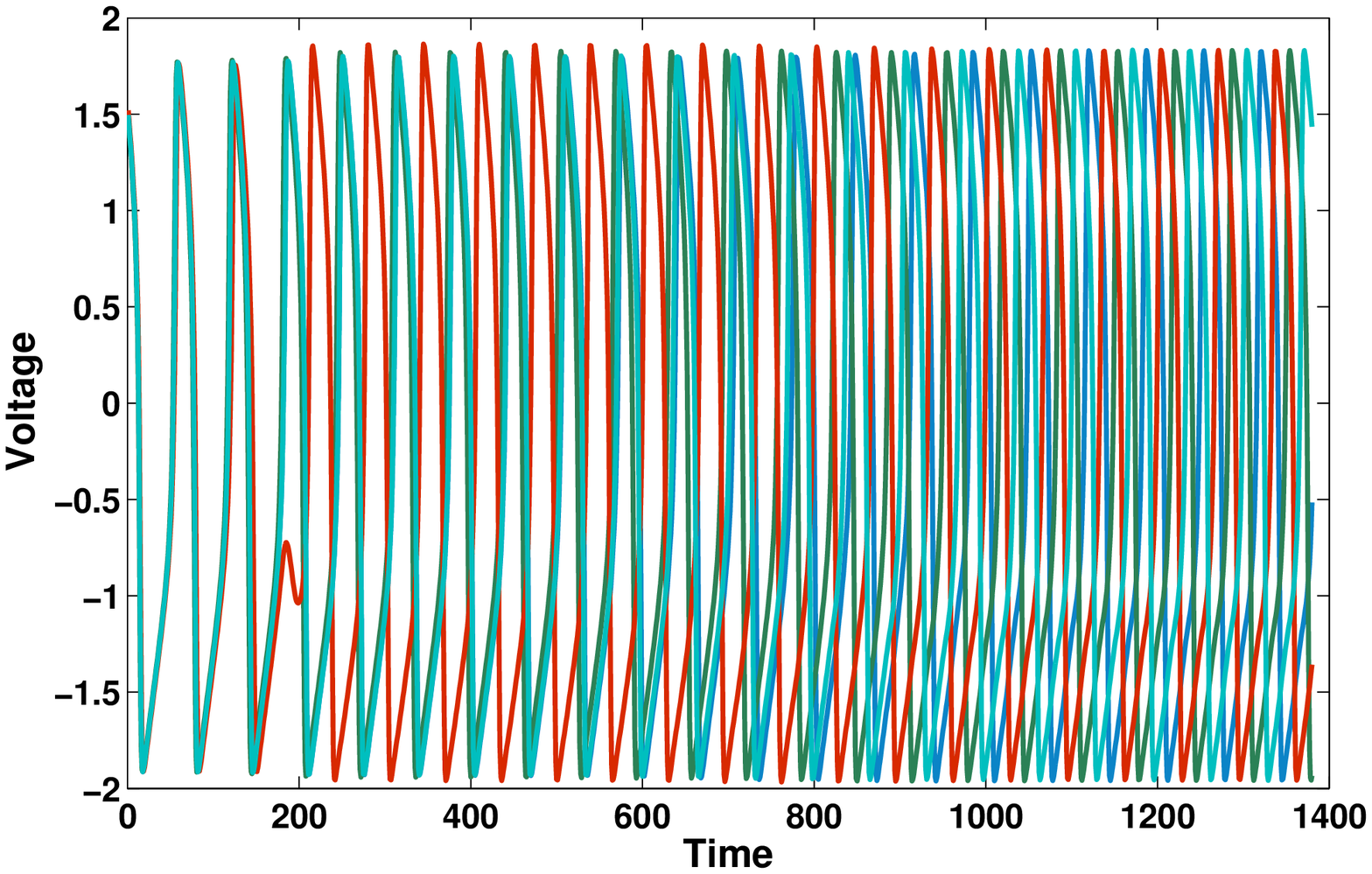}
\caption{\label{fig:FN3Desync} {\bf{Phase Shifted Synchronization in 3 and 4 Coupled Fitzhugh Nagumo Equations:}}(Parameter values as in \ref{fig:FN2Sync}).  From identical initial conditions the oscillators quickly desynchronize - the resulting dynamics gives spikes evenly distributed within each period.}
\end{center}
\end{figure}

\begin{figure}
\begin{center}
\includegraphics*[height=40mm, width=70mm]{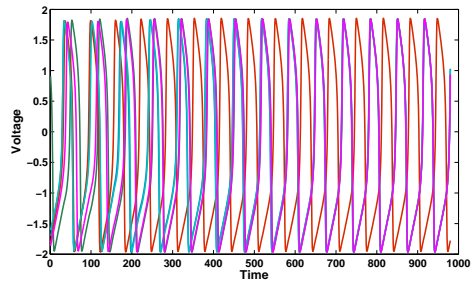}
\caption{\label{fig:FN5Desync} {\bf{Phase Shifted Synchronization (Clustering) in 5 Coupled Fitzhugh-Nagumo Equations:}}(Parameter values as in \ref{fig:FN2Sync}).  From non identical initial conditions the oscillators form into a cluster of 2 and a cluster of 3.  The clusters are evenly distributed around the phase.}
\end{center}
\end{figure}

It is straightforward using Equation \ref{eqn:FHN2}, to cause, for instance a population of $5$ neurons similar to Figure \ref{fig:FN5Desync} to exhibit all possible clusterings.\\

The number of possible arrangements of such clusters is given by the number of \textit{set partitions} on those neurons.  For instance: A very small brain consisting of $3$ neurons may possibly organize into all $3$ neurons acting in unison, $2$ neurons acting in unison and $1$ acting independently etc. The set of all $5$ possible cluster formations can be summarized as:

\begin{eqnarray*}
& &\{1, 2, 3\}\\ \nonumber
& &\{ \{1, 2\}, \, \{3\} \}\\ \nonumber
& &\{ \{1, 3\}, \, \{2\} \}\\ \nonumber
& &\{ \{2, 3\}, \, \{1\} \}\\ \nonumber
& &\{ \{1\}, \, \{2\}, \, \{3\}\}. \nonumber
\end{eqnarray*}

The enumeration of the set partitions for a given number of objects (n) is given by the Bell number B(n).  Bell numbers can be calculated using the recurrence relation:

\begin{equation}
B_n=\sum_{k=0}^{n-1}B_k 
\left( \begin{array}{c} 
n-1 \\ 
k
\end{array}\right). 
\end{equation}

The first 10 bell numbers are:  1, 1, 2, 5, 15, 52, 203, 877, 4140,   21147, 115975, 678570.\\

B(15) is $190899322$.  This is only bell(bell(5)).\\

As Bell numbers increase very rapidly it is not possible to directly calculate the exact number of set partitions for $n=8.6\times10^{12}$. Computational requirements make this unfeasible.  We can approximate large Bell numbers using the asymptotic approximation \cite{Lovasz}:

\begin{equation}
B(n) \sim \frac{1}{\sqrt{n}}\left(\frac{n}{W(n)}\right)^{n+\frac{1}{2}}e^{\left({\frac{n}{W(n)}-n-1}\right),}
\end{equation}
where $W(n)$ is the Lambert W function.\\

This gives as an initial upper bound based solely on the number of set partitions (and therefore synchronization cluster states) as:

\begin{equation}
B(8.6\times 10^{12}) \sim 5.927 \times 10^{95,401,985,845,526}.
\end{equation}

Which is considerably larger than the previous estimate in Science and massively larger than the total computing power of the world to date.  Score one for evolution!\\

\section{Ordering of Cluster States}
So far we have only considered the number of possible cluster states that $8.6\times10^{12}$ could exist in.  We are only considering set partitions which are not ordered.  For instance $$\{ \{1, 2\}, \, \{3\}\} = \{\{3\}, \{1,2\} \},$$ but what if the temporal order in which each cluster of neurons fired was also a part of the coding mechanism.  We now have to consider the number of {\em{permutations}} of the possible cluster states.\\

For a given set $n$ the number of permutations $P(n)=n!$ - however it would be an oversimplification to just take the factorial of the Bell number calculated above.\\

Consider a brain in which the neurons had formed into $3$ clusters and the order in which the clusters fire is relevant.  The possible orderings are:
\begin{eqnarray*}
{1,2,3}\\
{1,3,2}\\
{2,1,3}\\
{2,3,1}\\
{3,1,2}\\
{3,2,1}
\end{eqnarray*}
but from a coding perspective many of these would be equivalent.  For instance, if we take the first permutation and imagine the neuron clusters repeatedly firing in this order, we would have the firing pattern ${1,2,3,1,2,3,1,2,3,\dots}$ which would be the same ordering as if we took the second or fourth example above - we need to discount any {\em{cyclic permutations}} of orderings which we have already considered. The formula for this is given as  \cite{Weisstein} 
\begin{equation}
P(n)=(n-1)!
\end{equation}\\

If, as previously explained, we allow for all possible cluster states, a brain of $n$ neurons could form into any number of clusters $n_c \in {1,\, 2,\, \dots,\, n}$ where $n_c=n$ would be the completely desynchronized state and $n_c=1$ would be completely synchronized (neither of which would be particularly healthy).\\

We are therefore required to compute 
\begin{equation}
\sum_{k=1}^n(k-1)! \textrm{ for } n=5.927 \times 10^{95,401,985,845,526}
\end{equation}.

We can approximate the factorial sum using the expansion 
\begin{equation}
\sum_{k=1}^nk! \sim n!\left(1+\frac{1}{n}+\frac{1}{n^2}+\frac{2}{n^3}+\frac{5}{n^4}+\frac{15}{n^5}+\mathcal{O}\frac{1}{n^5}\right)
\end{equation}
which can be derived from Stirling's formula \cite{abramowitz}.\\

Again it is not possible to directly determine $(n-1)!$ for such a large number but we can approximate the factorial using the asymptotic formula of Ramanujan \cite{Karatsuba}  which gives:
\begin{equation} \label{eqn:Ramanujan}
    n! \sim \sqrt{\pi}\left(\frac{n}{e}\right)^n \sqrt[6]{8n^3+4n^2+n+\frac{1}{30}}.
\end{equation}

Clearly for such a large $n$ the $4n^2+n+\frac{1}{30}$ terms in the Eqn. \ref{eqn:Ramanujan} are significantly smaller than the $n^3$ terms and as such will not be considered.  Taking logarithms of Eqn. \ref{eqn:Ramanujan} gives us the approximation:
\begin{eqnarray}
\log(n!) & \sim &\log(\sqrt{\pi}\left(\frac{n}{e}\right)^n \sqrt[6]{8n^3}) \nonumber\\
         & \sim & \log\sqrt{\pi}+n\log\left(\frac{n}{e}\right)+\log\sqrt[6]{8n^3} \nonumber\\
         & \sim & n\log\left(\frac{n}{e}\right)
\end{eqnarray}
if we consider only the highest order terms.\\

This gives us a final estimate of the upper bound on the number of computational states for the human brain to be of the order:\\
{\Large{$$10^{565447570106432 \times 10^{95,401,985,845,526}}$$}}\\
which is considerably larger than the total number $2^n$ of binary states possible \footnote{where $n$ is the number of transistors on the planet today} if every computer, mobile phone, pocket calculator and wi-fi enable refrigerator, ever built, were wired together into one giant energy sucking super-machine. Yet we power a human brain for an hour on the calorific content of one apple. \footnote{The average human brain uses about 20 Watts \cite{Drubach}, the total power of all computing devices is in the order of many petaWatts ie $>10^{15}$ }

\bibliographystyle{plain} 
\bibliography{BSBib}

\end{document}